# Satellite Constellations Exceed the Limits of Acceptable Brightness Established by the IAU

Anthony Mallama[1] and Richard E. Cole[1]

2025 June 30

[1] IAU - Centre for the Protection of the Dark and Quiet Sky from Satellite Constellation Interference

Correspondence: anthony.mallama@gmail.com

Abstract

Brightness statistics for satellites of the Starlink, BlueBird, Qianfan, Guowang and OneWeb constellations are reported. The means and standard deviations are compared to acceptable limits set by the International Astronomical Union's Centre for the Protection of the Dark and Quiet Sky From Satellite Constellations Interference. Nearly all these spacecraft exceed the magnitude 7+ brightness limit pertaining to interference with professional research. Most also exceed the magnitude 6 reference where they distract from aesthetic appreciation of the night sky.

## 1. Introduction

Bright spacecraft interfere with astronomical research (Barentine et al. 2023) and spoil aesthetic appreciation of the night sky (Mallama and Young 2021). In response to this problem, the International Astronomical Union created a Centre for the Protection of the Dark and Quiet Sky from Satellite Constellation Interference (IAU CPS). This paper reports on a comparison between observed magnitudes of constellation satellites and brightness limits established by the IAU CPS.

## 2. Magnitude limits

The IAU CPS recently issued a call to protect dark and radio quiet skies (IAU 2024 and Boley et al. 2025). This report included a recommendation that satellites in operational orbits should never be visible to the unaided eye. Objects of visual magnitude 6 can be seen at locations where the sky is minimally affected by light pollution. We refer to this magnitude as the *aesthetic reference*.

The IAU CPS statement also defined a maximum brightness for interference with professional astronomy which we call the *research limit*. For altitudes up to 550 km that limit is magnitude 7.0. Equation 1 specifies the limit for altitudes above 550 km,

$$M_V > 7.0 + 2.5 * \log_{10}(\text{altitude} / 550)$$

Equation 1

where $M_V$ is the Johnson visual magnitude and *altitude* refers to the satellite's height in km above sea level.

### 3. Observed magnitudes

Magnitude statistics taken from comprehensive studies of individual constellations are listed in Table 1. Most of these results have already been published as indicated in the 'Source' column; 'plus' in that column indicates that new observations have been added. Values for Guowang, Starlink V1.5 and Starlink Mini satellites at 450 and 485 km have not previously been published and were derived for this study.

The statistics combine electronic $M_V$ magnitudes with others obtained by eye. The spectral sensitivities of these bands are nearly the same, so they are combined.

Observations were recorded when the satellites were at their operational heights for all constellations except Qianfan and Guowang. The Qianfan spacecraft are currently orbit-raising from initial altitudes of 800 km and higher to about 1,070, and their mean observed height was 955. Likewise, the Guowang satellites are rising from 900 km and higher to about 1,170, with a mean of 1,053.

### 4. Comparison with IAU CPS limits

The mean apparent magnitudes and standard deviations for satellites presently being launched are listed in the top section of Table 1. Those statistics are plotted as a function of altitude in Figure 1.

The mean apparent brightness for every satellite constellation except OneWeb exceeds the *research limit*. Approximately half of OneWeb magnitudes also exceed that limit. For other constellations, one standard deviation fainter than the mean still exceeds the *research limit*.

With regard to the *aesthetic reference*, all the mean values are brighter except for OneWeb and Starlink Mini satellites at 485 km. One standard deviation brighter than the mean for the 485 km Minis also exceeds the *aesthetic reference*. The bright extreme of the distribution for OneWeb does not exceed that limit.

Besides the satellites discussed above, SpaceX launched three models of Starlink spacecraft that are now discontinued. Furthermore, their Mini satellites are no longer orbit-raising to 550 km. These four cases are listed in the bottom section of the Table. All the mean apparent brightness values exceed the *research limit*, while those of V1.0 and V1.5 also exceed the *aesthetic reference*.

### 5. Discussion

Table 1 also reports statistics for apparent magnitudes that are adjusted to a range of 1,000 km. This allows for comparison between different constellations at a uniform distance, an indication of efficiency of brightness mitigation.

The BlueBird constellation is the brightest by apparent and 1000-km magnitude. However, their numbers are far smaller than the other constellations studied here.

The mean 1000-km brightness of Starlink Mini satellites orbiting at 550 km from Generation 2 are fainter than all models of Starlink Gen 1 spacecraft. This illustrates the successful brightness mitigation implemented by SpaceX, because the Minis are more than four times as large. However, the mean apparent brightness of the subsequent lower altitude Gen 2 Mini spacecraft (DTC, 450 km and 485 km) are greater than the Minis at 550 km due to their smaller distances.

Table 1. Satellite Magnitude Statistics

```
Currently launching:
Constellation    Height   --- Apparent ---    --- 1000-km ---     Mags       Source
-------------    ------   Mean  StDv  SdMn    Mean  StDv  SdMn    ------     ----------
SL-Mini-DTC       350     5.16  1.30  0.06    6.47  1.32  0.06      549      Mallama et al 2025a
SL-Mini-450       450     5.97  0.74  0.03    7.15  0.63  0.03      534      This paper
SL-Mini-485       485     6.24  0.75  0.03    7.15  0.70  0.03      570      This paper
BlueBirds         500     3.30  1.70  0.07    3.77  1.45  0.06      546      Cole et al 2025 plus
Qianfan           955     5.76  0.94  0.04    5.24  0.72  0.04    1,161      Mallama et al 2025b
Guowang          1053     5.07  1.16  0.05    4.21  0.85  0.04      455      This paper
OneWeb           1200     7.85  0.71  0.01    7.05  0.66  0.01   80,000      Mallama 2022

Discontinued:
Constellation    Height   --- Apparent ---    --- 1000-km ---     Mags       Source
-------------    ------   Mean  STDv  SDMn    Mean  STDv  SDMn    ------     ----------
SL-V1.0           550     5.05  0.52  0.00    5.89  0.46  0.00   60,000      Mallama 2021
SL-VisorSat       550     6.43  0.86  0.00    7.21  0.89  0.00   40,000      Mallama 2021
SL-V1.5           550     5.67  0.73  0.04    6.34  0.86  0.05      268      This paper
SL-Mini-550       550     6.36  0.63  0.01    7.22  0.83  0.01    3,869      Mallama et al 2023 plus
```

'SL' - Starlink, 'StDv' - standard deviation, 'SdMn' - standard deviation of the mean

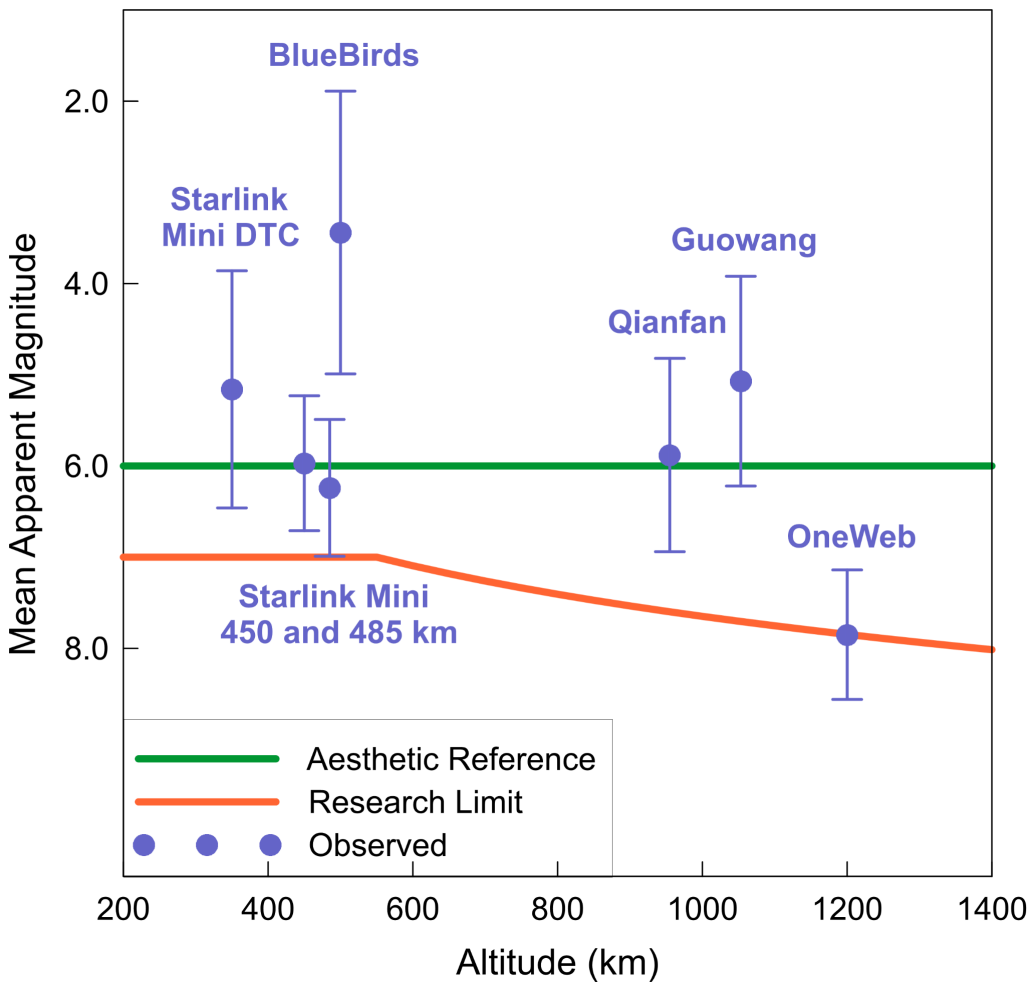

Figure 1. Brightness means and standard deviations of satellite constellations by height.

## 6. Conclusions

Brightness statistics for satellites of the Starlink, BlueBird, Qianfan, Guowang and OneWeb constellations are reported. The means and standard deviations are compared to acceptable limits set by the IAU CPS. Nearly all these spacecraft exceed the magnitude 7+ brightness limit which pertains to interference with professional research. Most also exceed the magnitude 6 reference where they distract from aesthetic appreciation of the night sky.

The BlueBird constellation is the brightest but also the smallest in number. SpaceX has successfully applied brightness mitigation to their Starlink spacecraft. However, that dimming is partially offset by the lower altitudes of the satellites currently being launched.

## Acknowledgments

We thank E. Katkova for maintaining the database of [MMT9](#) magnitudes online. J. McDowell's [graphs](#) of satellite altitude versus time were important for this study. The [Heaven-Above](#) website was used to predict satellite passes. The planetarium program, [Stellarium](#), and the [Orbitron](#) app were used for data analysis.